# Determinantes do planejamento estratégico da rede de uma companhia aérea

Bruno Felipe de Oliveira
Alessandro V. M. Oliveira↗
Instituto Tecnológico de Aeronáutica, São José dos Campos, Brasil
↗ Autor correspondente. Instituto Tecnológico de Aeronáutica. Praça Marechal Eduardo Gomes, 50. 12.280-250 - São José dos Campos, SP - Brasil.
E-mail: alessandro@ita.br.

*Resumo*: Este trabalho tem como foco tentar entender como é feita a construção de malha de uma companhia aérea. Para isso foi estudado o caso da Azul, investigando quais e como os fatores afetam a decisão de entrada dessa companhia aérea em rotas domésticas, além de analisar como a fusão da Azul com a companhia aérea regional Trip afetou o planejamento de malha da companhia. Para isso, foi feito um estudo acadêmico usando um modelo econométrico para entender o modelo de entrada da companhia aérea. Os resultados mostram que o modelo de negócios da Azul é baseado em conectar novos destinos, ainda não servidos por seus concorrentes, a um de seus hubs, e consistentemente evitar rotas ou aeroportos dominados por outras companhias aéreas. Em relação aos efeitos da fusão, os resultados sugerem que a Azul se afastou do seu modelo de entrada original, baseado na JetBlue, para um modelo mais orientado à aviação regional, entrando em rotas mais curtas e em aeroportos regionais.

*Palavras-chave*: transporte aéreo, companhias aéreas, econometria.

## I. INTRODUÇÃO

A construção de malha das companhias aéreas é uma das áreas mais estudadas no campo da aviação civil. A entrada no mercado, por si só, é um importante tema empírico na literatura de economia, estratégia e marketing (Dixit & Chintagunta, 2007), e é altamente importante para a competição e a inovação em qualquer campo (Hüschelrath & Müller, 2011).

Um dos principais acontecimentos que afetou a competitividade e a inovação no campo da aviação civil foi a Lei de Desregulamentação dos EUA em 1978. Essa foi uma lei americana que desregulamentou o setor de aviação naquele país, removendo o controle do governo sobre tarifas, rotas e entrada de novas companhias aéreas, introduzindo efetivamente um mercado livre no setor da aviação civil americana. Um grande número de novas companhias aéreas ingressou no setor, enquanto as companhias aéreas existentes expandiram suas operações para novos mercados. Os preços das passagens aéreas caíram quando as empresas aéreas aumentaram sua produtividade, repassando essas economias aos consumidores finais. Além disso, houve a expansão de companhias aéreas de baixo custo, que serviu para reduzir ainda mais os preços das passagens aéreas (Dresner et al. 2002).

A desregulamentação do mercado de aviação civil americana e seus efeitos na indústria inspiraram outras regiões do mundo a liberalizar seus próprios mercados, como a Europa, entre 1987 e 1997, que permitiu às companhias aéreas daquela região a operar voos domésticos nos países membros (Gil-Moltó & Piga, 2008) e Brasil, entre 1989 e 2001, que implantou o regime de liberdade tarifária no transporte aéreo doméstico.

A literatura nesta área abrange amplamente os benefícios da entrada de companhias aéreas no mercado, dizendo que essa entrada aumenta as pressões competitivas, obrigando as empresas aéreas a aumentar sua eficiência produtiva, consequentemente levando a uma redução no preço das passagens e a melhora no nível de serviço (Hüschelrath & Müller, 2011). Esses benefícios atraem um interesse considerável dos participantes desse setor. Os passageiros desse sistema, por exemplo, são atraídos pela ideia de que a entrada de uma nova companhia aérea pode diminuir os preços das passagens e aumentar o nível de serviço do voo.

Por outro lado, o governo local, regional ou nacional pode estar interessado na atratividade de um determinado aeroporto para as companhias aéreas do setor, a fim de avaliar a necessidade de investimentos para atender à demanda que poderá surgir devido à entrada de uma companhia aérea, permitindo um melhor planejamento das despesas do governo. Neste contexto, este trabalho tentará explicar ao leitor como as companhias aéreas decidem seus destinos usando o caso da companhia Azul Linhas Aéreas no mercado da aviação doméstica brasileira.

A Azul Linhas Aéreas é uma companhia low cost fundada em 2008 pelo empresário David Neeleman, fundador das low costs JetBlue e WestJet, e que serviriam de modelo para a Azul. Neeleman fundou essa empresa companhia aérea brasileira no contexto da expansão da low cost Gol no início dos anos 2000 (Oliveira, 2008), e especialmente após a declaração de falência da Varig - uma das maiores companhias aéreas brasileira na época, cuja saída do mercado deixaria espaços a serem preenchidos (de Oliveira, 2017).

Neste cenário, a Azul iniciou suas operações em dezembro de 2008, servindo apenas três destinos: Viracopos, Salvador e Porto Alegre, sendo que o aeroporto de Viracopos foi indicado como hub da Azul desde o início de suas operações. Hoje, a Azul é uma das maiores companhias aéreas do Brasil, com uma frota de 134 aeronaves, 104 destinos e 30% do número total de voos no país . A Figura 1 mostra a evolução da malha aérea da Azul no Brasil.

Além do crescimento da Azul no território nacional, também é possível verificar o crescimento dessa companhia aérea ao nível aeroportuário, conforme a Figura 2, que mostra um gráfico com a evolução na movimentação anual de passageiros no aeroporto de Viracopos. Sendo um aeroporto secundário, localizado próximo aos dois maiores aeroportos do Brasil (Guarulhos e Congonhas), Viracopos historicamente apresentava um movimento anual de passageiros abaixo de 0,9 milhão. Após a entrada da Azul no mercado brasileiro em 2008 e a escolha de Viracopos como seu hub, o movimento anual de passageiros apresentou um crescimento impressionante, chegando a 10 milhões em 2014.

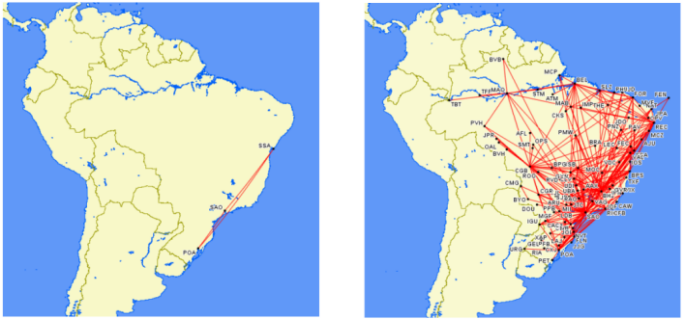

Fonte: VRA/ANAC, elaborado em www.gcmap.com.
**Figura 1 – Malha aérea da Azul - 2008 (esq.) e 2018 (dir.)**

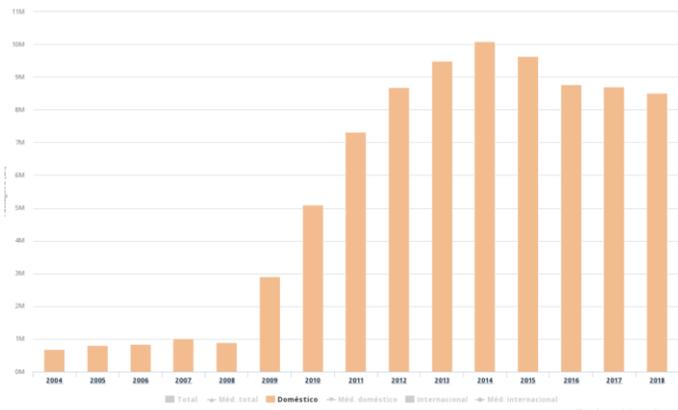

Fonte: Hórus Labtrans (2019).
**Figura 2 – Evolução da Azul no aeroporto de Viracopos**

Para continuar crescendo no mercado brasileiro, a Azul anunciou em 2018 sua intenção de entrar em 25 destinos domésticos nos próximos anos . A partir de uma investigação dessas cidades, verificou-se que maioria delas está localizada nas regiões Sul, Sudeste e Nordeste do Brasil. Desta forma, um dos objetivos deste trabalho é entender como a Azul escolhe seus novos destinos domésticos e quais são características de mercado mais importantes para ela. Acredita-se que, ao entender os critérios de seleção dos destinos por parte das companhias aéreas, este estudo trará benefícios ao planejamento de investimentos nos aeroportos regionais e a expansão da aviação civil no Brasil.

Além de investigar o modelo de entrada da Azul em rotas domésticas, este trabalho também visa entender os efeitos que uma fusão pode ter no modelo de entrada de companhias aéreas. Em 2012, a Azul e a Trip anunciaram a fusão de suas operações , sendo que a Trip era uma companhia aérea brasileira e a maior do segmento regional na América Latina. Essa fusão fez com que o grupo Azul-Trip se tornasse a terceira maior companhia aérea do Brasil, atendendo 96 destinos domésticos . Com a rápida expansão da Azul no mercado brasileiro, especialmente após a fusão com a regional Trip, esse trabalho também tenta entender como essa fusão pode ter afetado o planejamento de malha aérea e o modelo de negócios da Azul. Para isso, é necessário entender como funciona o planejamento da malha aérea das companhias.

## II. Planejamento estratégico de rede

Quando uma companhia está planejando a sua malha aérea, existem certas características de mercado que são colocadas à mesa para uma análise prévia da viabilidade de uma rota.

Uma das principais características analisadas pelas companhias aéreas é a existência de demanda por voos devido a uma alta atividade econômica naquele local. Essa característica pode ser medida pela população local, PIB per capita, quantidade de turistas ou mesmo o tráfego de passageiros, se houver tráfego aéreo no local.

A geografia também é levada em consideração no planejamento de uma rota. Um exemplo é proximidade entre aeroportos. Quando dois aeroportos, um grande e outro pequeno em relação ao volume de passageiros, se localizam próximos numa área, a tendência é que o aeroporto grande capture a demanda do outro, mesmo que o aeroporto pequeno se localize em uma cidade com alta atividade econômica. Isso acontece porque os potenciais usuários do segundo aeroporto enxergam uma maior flexibilidade na escolha e compra de passagens aéreas no aeroporto grande, que possui um maior volume de voos.

Outro fator geográfico considerado pelas companhias aéreas é a centralidade do aeroporto em relação aos grandes centros urbanos ou regionais. Esse fator é muito importante para as companhias aéreas, pois elas avaliam a possibilidade da criação de hubs operacionais que possam gerar voos de conexão com outros aeroportos próximos da região. Um exemplo disso são os hubs da própria Azul, que concentra as suas operações na região nordeste no aeroporto de Recife, na região centro-oeste no aeroporto de Cuiabá, e em Viracopos na região sudeste. De forma similar, é considerada também a centralidade do aeroporto em relação à área do país. Um exemplo disso é o aeroporto de Brasília, que é o hub operacional da companhia aérea LATAM.

Os fatores citados anteriormente são levados em consideração principalmente quando uma companhia aérea está entrando em uma cidade que não está oferecendo voos, ou seja, novos destinos. Porém, não é raro que as companhias aéreas entrem em cidades já servidas por voos regulares. Para esses casos, as companhias aéreas tentam entender como funcionam esses aeroportos: se ele é hub de uma companhia aérea concorrente, ou se o aeroporto está congestionado e está suscetível (ou já possui) a restrição de slots aeroportuários.

No tema da competição entre as companhias aéreas, elas tentam evitar a concorrência sempre que possível, com exceção das rotas-tronco, que são bastante competitivas devido a sua alta demanda de voos e lucratividade. Neste caso, as companhias aéreas entrantes verificam a concentração de mercado daquela rota ou aeroporto, ou seja, se ela é dominada por uma companhia aérea. Além disso, uma companhia aérea entrante também pode verificar quais as empresas já operam naquele mercado, para verificar a similaridade do modelo de negócio entre as empresas e avaliar se há algum tipo de demanda não sendo atendida atualmente.

Todos esses fatores têm que estar alinhados com o modelo de negócios da empresa. Uma companhia aérea regional irá servir primariamente rotas regionais, ligando vários aeroportos pequenos a um grande centro urbano. Isso pode refletir na etapa média de voos dessa companhia aérea, que será menor se comparada às companhias aéreas que ligam apenas os grandes centros urbanos.

Paralelamente, as características da frota de aeronaves de uma companhia aérea também afeta as características das rotas servidas por ela. Por exemplo, uma companhia que possui aeronaves menores com baixa autonomia nos voos dificilmente operará rotas longas; ou então pode não ser vantajoso para uma companhia aérea utilizar aeronaves grandes em rotas com baixa densidade de passageiros, já que o fator de ocupação dos assentos poderá ficar abaixo do ponto de equilíbrio para justificar, financeiramente, a operação. Neste contexto, uma companhia aérea que possui uma maior diversidade de frota terá uma maior flexibilidade para operar diferentes destinos.

O modelo de negócios pode afetar, ainda, o tipo da malha a ser construída pela companhia: se ela será baseada num sistema hub-and-spoke, com conexões nos aeroportos centrais, ou se a malha será baseada em voos ponto-a-ponto. Porém, vale notar que conforme a companhia vai crescendo, é comum seu modelo de negócio se tornar cada vez mais difuso, visto que ela vai entrando em todos os mercados que qualquer companhia aérea grande entraria.

### III. Revisão da Literatura

Desde a desregulamentação da aviação civil americana, os pesquisadores têm feito pesquisas acerca da competição no mercado do transporte aéreo nos temas do preço das passagens aéreas e do modelo de entrada em rotas. Diversos estudos foram feitos nessas áreas após este evento nos EUA e após outros mercados passarem pelo processo de liberalização também, como foi o caso da Europa e em alguns países da América do Sul e Ásia no período entre 1987 e 2007 (Assaf e Gillen, 2012), mostrando que é um tema relevante a ser estudado. Desta forma, para servir de base para o estudo da Azul, esta seção apresentará os estudos da literatura em que os pesquisadores procuraram entender o modelo de entrada das companhias aéreas, mostrando quais as características de mercado relevantes para a escolha de um destino ou rota.

Um trabalho clássico é de Morrison e Winston (1990), da Northeastern University e The Brookings Institution, em que os autores analisaram a dinâmica de preços e concorrência de companhias aéreas na indústria. Com relação à entrada e saída de companhias aéreas, eles usaram um modelo estatístico de probabilidade de entrada e descobriram que, quando uma companhia aérea já está operando em um par de aeroportos, existe uma grande probabilidade de esta companhia criar uma rota entre esses dois aeroportos.

Sinclair (1995), da J.W. Wilson Associates, expandiu a literatura da área ao mostrar fortes evidências de que a decisão de entrada e saída pelas companhias aéreas é afetada pelo tamanho e pela utilização de um sistema hub-and-spoke. Segundo esse autor, a companhia com um forte sistema de hub pode inibir a entrada de concorrentes, enquanto um entrante com um forte sistema de hub entrará mais fácil nas rotas.

Dresner, Windle e Yao (2002), da University of Maryland, estudaram os efeitos das barreiras de entrada, mostrando que os controles de slots, restrições de gates e a utilização destes durante o horário de pico afetam negativamente a decisão de entrada em um par de aeroportos. Gil-Moltó e Piga (2008), da University of Leicester e Loughborough University, analisaram o mercado aéreo europeu em relação à entrada de companhias aéreas tradicionais e low costs carriers (LCCs). Entre as diferentes variáveis testadas, algumas delas confirmaram o que foi observado em estudos anteriores, como a presença num par de cidades influenciando positivamente na criação de uma rota entre elas, mas algumas variáveis apresentaram resultados interessantes. Por exemplo, o número de empresas que já operam em uma rota está positivamente correlacionado à entrada. Isso foi explicado pelos autores pelo fato de que um número menor de empresas em uma rota se deve à presença de uma companhia aérea dominante e/ou de presença de barreiras de entrada nessa rota. Alternativamente, o tamanho do tráfego aéreo apresentou uma correlação negativa, que também pode ser explicada pela presença de uma companhia aérea dominante ou de barreiras de entrada.

Pesquisadores que também investigaram os casos específicos das low costs. Ito e Lee (2003), da Brown University e Compass Lexecon, analisaram o crescimento das empresas aéreas nos EUA e os fatores que influenciam sua entrada. De acordo com o trabalho deles, o fator mais impactante de uma entrada de LCC é a densidade do mercado. Boguslaski, Ito e Lee (2004), o primeiro da Gordian Group e Ito e Lee os mesmos autores do trabalho anterior, expandiram a literatura analisando a evolução das estratégias de entrada da Southwest, a maior LCC americana, ao longo dos anos. Eles encontraram uma mudança de comportamento dessa companhia aérea na escolha de rotas para operar, que inicialmente atuava em mercados densos e de curta distância, mas que passou a entrar em mercados de baixa densidade e longa distância. A conclusão que os autores tiveram é que as LCCs não são obrigadas a voar em mercados densos e de curta distância, focados em passageiros viajando a lazer, para fazer o seu modelo de negócio funcionar, e que, portanto, as companhias aéreas tradicionais estarão cada vez mais expostas à concorrência das low costs.

Oliveira (2008), do Instituto Tecnológico de Aeronáutica, analisou o padrão de entrada da LCC Gol no Brasil, e concluiu que o comportamento de entrada dessa companhia aérea era consistente com o padrão clássico de entrada da Southwest, focando em rotas densas e de curta distância. Além disso, o autor também encontrou evidências de que a Gol havia mudado seu padrão de entrada nos anos seguintes de sua fundação, ficando mais parecido com o padrão da JetBlue, que foca em rotas de longa distância. Essa mudança de comportamento foi explicada pelo autor como um efeito de idiossincrático do país, como economias de escopo não observadas no estudo.

Müller, Hüschelrath e Bilotkach (2012), da ZEW Centre for European Economic Research, WHU Otto Beisheim School of Management e Northumbria University, estudaram o padrão de entrada da LCC JetBlue Airlines na indústria de voos domésticos dos EUA. Eles mostraram que a JetBlue consistentemente evita aeroportos dominados por outras companhias aéreas; mas que ela atua em rotas dominadas, utilizando aeroportos secundários em rotas mais densas para ganhar uma vantagem no preço de suas passagens aéreas em relação às companhias aéreas tradicionais. Os autores também mostraram que uma das principais características da JetBlue é a sua presença em rotas de longa distância, além de evitar a concorrência direta com outras LCCs.

Boguslaski, Ito e Lee (2004) e Oliveira (2008) concluíram que os padrões de entrada das LCCs estavam mudando ao longo do tempo, e esse tema também foi estudado por de Wit e Zuidberg (2012), da University of Amsterdam e SEO Economic Research, em uma pesquisa sobre os limites de crescimento do modelo low cost. Eles analisaram o mercado aéreo europeu e americano, e concluíram dizendo que esses mercados haviam atingido uma alta maturidade, o que obrigaria as LCCs a adotarem novas estratégias em seus modelos de negócios, como a diminuição na frequência de voos e a atuação em rotas de longa distância. Eles também apontaram outras estratégias que poderiam ser adotadas pelas LCCs, como a entrada em aeroportos primários, a criação de hubs, criação de alianças ou acordos de codeshare com as companhias aéreas tradicionais, e aquisição ou fusão com outras companhias aéreas.

Dobruszkes, Givoni e Vowles (2017), da Université Libre de Bruxelles, Tel-Aviv University e University of Northern Colorado, estudaram a entrada das LCCs nos aeroportos primários, que foi uma das estratégias citada por de Wit e Zuidberg (2012). Os autores analisaram os mercados americano e europeu, e constataram que tanto a Southwest quanto a Ryanair, as maiores LCCs de seus respectivos mercados, apresentavam uma tendência de entrada nos principais aeroportos do país ou continente, o que confirma a mudança no comportamento de entrada das LCCs, mostrando que as companhias aéreas tradicionais estão cada vez mais suscetíveis à competição das low costs.

Como mostrado, as pesquisas mais recentes têm se preocupado em analisar a mudança no comportamento das LCCs em mercados maduros, como EUA e Europa, além de analisar os fatores que levam as companhias aéreas a entrar nas rotas nos mercados emergentes, mostrando que esse tema ainda é relevante para os pesquisadores atuais. Apesar de ser citado por de Wit e Zuidberg (2012), o tema da fusão e aquisição de companhias aéreas, e como isso impacta no seu planejamento de entrada em rotas é um tema não discutido pelos pesquisadores ainda. Desta forma, para preencher essa lacuna na literatura, foi feito um estudo nacional cujo objetivo foi identificar os fatores que influenciam na decisão de entrada da Azul em rotas domésticas brasileiras, e ainda verificar se houve uma mudança no planejamento de malha da Azul após sua fusão com a companhia aérea Trip.

## IV. Caso da Azul Linhas Aéreas

O estudo de Oliveira e Oliveira (2019) busca entender como a Azul escolhe seus novos destinos domésticos e quais as características de mercado mais importantes para ela. Esse modelo foi baseado na literatura científica existente acerca do tema, e visa trazer maior conhecimento nessa área, que pode ser usado para um melhor planejamento de investimentos nos aeroportos regionais e a expansão da aviação no Brasil. Para esse estudo, foi analisado o mercado do transporte aéreo doméstico no Brasil entre dezembro de 2008 a dezembro de 2018. Para facilitar a análise dos resultados, as características de mercado foram separadas em quatro grupos: distância, demanda, aeroporto e concorrência.

Para o grupo da distância, no estudo foram testadas rotas de curta distância (menores que 500 km), curta-média distância (entre 500 e 1000 km), média-longa distância (entre 1000 e 2000 km) e longa distância (maior que 2000 km). Para o grupo da demanda, foi testado se a densidade (a quantidade de passageiros existente) da rota é um fator importante para a Azul. Além disso, testamos também se a Azul leva em consideração, durante seu planejamento de malha, a quantidade de passageiros em conexão e a quantidade de passageiros em voos turísticos em uma rota. Para o grupo do aeroporto, testaram-se casos em que o aeroporto é um hub de uma companhia aérea concorrente à Azul, e também qual o impacto de aeroporto saturado na decisão da Azul.

Por fim, para o grupo da concorrência, testou-se um conjunto de fatores relacionados à concentração de mercado das rotas e aeroportos, além da presença de companhias aéreas com modelos de negócios similares à Azul: LCC (Gol e Webjet) e regionais (Trip, regionais pequenas). Também testamos a presença da Avianca, que, apesar de ser uma companhia aérea tradicional, iniciou suas operações no Brasil como OceanAir, uma companhia aérea regional.

O resumo dos resultados se encontra na Tabela 1, apresentando os resultados: (+) para características favoráveis à entrada da Azul; (-) para características inibidoras à entrada; e (NS) para características não significantes para a decisão de entrada da Azul. Para entender o comportamento da Azul em diferentes períodos, testamos os cenários: (1) o período completo da base de dados; (2) apenas o período antes da fusão; e (3) apenas o período pós-fusão.

Em relação aos fatores de distância, os dados mostram que a Azul evita rotas longas, mas que nem sempre foi assim. Antes da fusão, a Azul mostrava uma preferência por rotas de média e longa distância, consistente com o modelo de negócios da JetBlue. Mas após a sua fusão com a Trip, seu modelo de entrada nas rotas mudou para um comportamento orientado para aviação regional, concentrando-se em rotas mais curtas. Os resultados também mostram que Azul não segue a tendência verificada na literatura dos padrões de entrada das LCCs, em que autores como Boguslaki, Ito e Lee (2004), Oliveira (2008) e de Wit e Zuidberg (2012) observaram que as LCCs estavam entrando em rotas cada vez mais longas.

**Tabela 2 – Determinantes da configuração de malha aérea da Azul**

| Características | (1) Período completo | (2) Antes da fusão | (3) Depois da fusão |
|---|---|---|---|
| 250-500 km | Caso base | Caso base | Caso base |
| 500-1000 km | - | NS | - |
| 1000-2000 km | - | + | - |
| 2000+ km | - | NS | - |
| Passageiros | + | + | + |
| Passageiros em conexão | - | - | NS |
| Passageiros em conexão - Hub Azul | + | + | + |
| Turismo | + | + | + |
| Hub grande | + | NS | + |
| Congestionamento | NS | + | - |
| HHI da rota | - | - | - |
| HHI da rota - Hub Azul | + | + | + |
| HHI do aeroporto | - | - | - |
| Presença da Gol | - | - | - |
| Presença da Webjet | + | + | - |
| Presença da Trip | + | NS | + |
| Presença de regionais pequenas | - | - | - |
| Presença da Avianca | - | - | - |

Fonte: Oliveira e Oliveira (2019). *NS = efeito não significante.

Em relação às características de demanda, algumas delas apresentaram resultados esperados. Por exemplo, uma rota com densidade de passageiros (em voos regulares) e turistas (voos fretados) atraem a atenção da Azul, concordando com estudos anteriores. Um fato interessante é que, apesar de sempre ser um fator positivo, a importância da densidade da rota foi caindo ao longo dos anos, confirmando a literatura (Boguslaki, Ito & Lee, 2004; Oliveira, 2008; de Wit & Zuidberg, 2012), e mostrando que Azul está entrando rotas menos densas quando comparadas ao período inicial de suas operações no Brasil. Em relação aos passageiros em conexão, os resultados mostram que a Azul evitava rotas com um elevado número de passageiros em conexão de outras companhias aéreas, mas, após a fusão, essa variável deixou de ser levada em consideração pela Azul. Por outro lado, em relação aos passageiros em conexão da Azul, os resultados mostram que o modelo de negócios da Azul sempre se baseou na conexão de seus novos destinos aos hubs existentes.

Em relação aos fatores aeroportuários, os resultados mostram que, antes da fusão, a Azul não levava em consideração se uma rota estava ligada a um hub concorrente, mas após a fusão a Azul passa a entrar nesses aeroportos. Uma explicação para isso é que esses hubs são aeroportos com uma alta movimentação de passageiros, o que a Azul vê como uma oportunidade para participar desse mercado denso e disputar com as outras companhias pela preferência dos passageiros.

Em relação ao fator de saturação dos aeroportos, os dados mostram uma mudança de paradigma da Azul. Antes da fusão, a Azul entrava em aeroportos congestionados, mas após a fusão começa a evitar esse tipo de aeroporto. O comportamento anterior à fusão pode ser explicado pela expansão inicial da Azul no mercado doméstico brasileiro, em que ela entrou em aeroportos congestionados, como em São Paulo e Rio de Janeiro, para estabelecer sua posição no mercado. Porém, após a fusão ela passa a evitar os aeroportos congestionados, pois a Azul já havia estabelecido sua posição no mercado doméstico, além dar maior foco em seu segmento regional.

E em relação às variáveis de competição, a maioria delas apresentou resultados esperados. Os dados mostraram que, tanto antes quanto após a fusão, a Azul evita rotas dominadas por outras companhias aéreas, a menos que a própria Azul possa ser a dominante. Por exemplo, a Azul tende a entrar numa nova rota quando existe a possibilidade de conectar esse novo destino a um de seus hubs, e ser a única companhia aérea atuando nessa rota. Analisando o nível de concentração de mercado no aeroporto, a Azul evita entrar em rotas que se conectam a aeroportos dominados por outra companhia aérea, embora seus efeitos negativos tenham diminuído após a fusão. Em relação à presença de outras companhias aéreas na rota, a Azul evita a concorrência direta com a Gol e as pequenas companhias aéreas regionais. Isso se justifica pela similaridade de seus modelos de negócios, fazendo com que a Azul não possa oferecer um produto diferenciado aos passageiros dessas companhias aéreas.

A partir desses resultados, é possível encontrar comportamentos consistentes da Azul ao longo do tempo, e que podemos chamar de seu modelo core business. A Azul tende a entrar em rotas densas, com passageiros regulares ou que viajam a turismo. Em relação aos rivais de mercado, a Azul evita competição direta com outras companhias aéreas quando estas já dominam as rotas e os aeroportos. Porém, uma característica principal da Azul é que ela tende a entrar em rotas em que ela pode ser a dominante, conectando esse novo destino com um de seus hubs.

Para validar esses resultados, nós investigamos alguns dos novos destinos da Azul. Neste trabalho foi falado que a Azul planejava entrar em 25 novos destinos domésticos a partir de 2019; desse grupo, verificamos que a Azul já entrou em três deles no momento da redação deste texto, que são: Mossoró-RN, Pato Branco-PR e Toledo-PR.

Esses três destinos compartilham a mesma característica de distância: todos estão conectados a um aeroporto a 500 km, mostrando o aspecto regional da Azul pós-fusão. Mossoró está conectado a Recife, enquanto Pato Branco e Toledo estão conectados a Curitiba. Além disso, o aeroporto Recife é um hub da Azul desde 2017, enquanto o aeroporto de Curitiba é considerado pela Azul como um "mini-hub". Além disso, outra característica similar entre esses aeroportos é que eles estavam ociosos até a entrada da Azul.

Essas evidências empíricas confirmam os resultados encontrados no estudo, mostrando que o modelo de entrada da Azul é baseado na conexão dos novos destinos a um de seus hubs ou aeroportos próximos, e que ela entra sozinha nessas cidades para dominá-las.

## V. Conclusões

E afinal, por que a minha cidade não tem voos? Como pôde ser visto neste trabalho, há muitos fatores que as companhias aéreas precisam levar em consideração no seu planejamento de malha aérea. Uma das principais características analisadas por elas é a existência de demanda por voos, ou seja, a atividade econômica de uma região. Se não há essa demanda, é muito difícil para o planejador de malha aérea justificar a presença da companhia aérea no local, a não ser que o aeroporto de sua cidade possua uma boa localidade geográfica para servir como um hub no futuro. Agora, se uma cidade ou região possui uma alta atividade econômica e mesmo assim não há voos para o aeroporto local, pode ser que existam outros fatores que impeçam a viabilidade das operações das companhias aéreas nesse aeroporto, como a proximidade com outros aeroportos grandes que tenha a preferência dos passageiros da região.

Mesmo com essas justificativas, pode ser que um dia que o aeroporto de sua cidade receba voos da Azul, que possui as suas próprias preferências como foi mostrado neste trabalho. O estudo em que este trabalho foi baseado investigou as características que a Azul busca em seus destinos e mostrou que, além de potencial de demanda do local, a Azul tem preferido entrar em rotas curtas, entre 250 a 500 km de distância, focando no seu segmento regional após a sua fusão com a Trip em 2012, e criando conexões com seus hubs existentes. Se o aeroporto já possui operações de outras companhias aéreas, é provável que a Azul não entre nesse mercado, já que ela evita sistematicamente aeroportos e rotas com a presença de rivais. Agora, se atualmente não há oferta de voos no aeroporto, isso pode ser um indicativo de que a Azul possa entrar nesta cidade no futuro, já que segundo o estudo ela prefere atuar sozinha numa rota, evitando a concorrência direta com seus rivais.

## Referências

Assaf, A. G., & Gillen, D. (2012). Measuring the joint impact of governance form and economic regulation on airport efficiency. European journal of operational research, 220(1), 187-198.

Boguslaski, C., Ito, H., & Lee, D. (2004). Entry patterns in the southwest airlines route system. Review of Industrial Organization, 25(3), 317-350.

de Wit, J. G., & Zuidberg, J. (2012). The growth limits of the low cost carrier model. Journal of Air Transport Management, 21, 17-23.

de Oliveira, R. P. (2017) Airline financial distress and its impacts on airfares: an econometric model of the dynamic effects of a bankruptcy filing followed by an acquisition. Thesis of Master in Science in Aeronautical Infrastructure Engineering, Field of Airports and Air Transportation – Instituto Tecnológico de Aeronáutica, São José dos Campos.

de Oliveira, R. P., & Oliveira, A. V. M. (2021). Financial distress, survival network design strategies, and airline pricing: An event study of a merger between a bankrupt FSC and an LCC in Brazil. Journal of Air Transport Management, 92, 102044.

Dixit, A., & Chintagunta, P. K. (2007). Learning and exit behavior of new entrant discount airlines from city-pair markets. Journal of Marketing, 71(2), 150-168.

Dobruszkes, F., Givoni, M., & Vowles, T. (2017). Hello major airports, goodbye regional airports? Recent changes in European and US low-cost airline airport choice. Journal of Air Transport Management, 59, 50-62.

Dresner, M., Windle, R., & Yao, Y. (2002). Airport Barriers to Entry in the US. Journal of Transport Economics and Policy (JTEP), 36(3), 389-405.

Gil-Moltó, M. J., & Piga, C. A. (2008). Entry and exit by European low-cost and traditional carriers. Tourism Economics, 14(3), 577-598.

Hüschelrath, K., & Müller, K. (2011). Patterns and effects of entry in US airline markets. Journal of Industry, Competition and Trade, 13(2), 221-253.

Ito, H., & Lee, D. (2003). Low cost carrier growth in the US airline industry: past, present, and future. Brown University Department of Economics Paper, (2003-12).

Morrison, S. A., & Winston, C. (1990). The dynamics of airline pricing and competition. The American Economic Review, 80(2), 389.

Müller, K., Hüschelrath, K., & Bilotkach, V. (2012). The Construction of a Low-Cost Airline Network–Facing Competition and Exploring New Markets. Managerial and decision economics, 33(7-8), 485-499.

Oliveira, A. V. M. (2008). An empirical model of low-cost carrier entry. Transportation Research Part A: Policy and Practice, 42(4), 673-695.

Oliveira, B. F., & Oliveira, A. V. M. (2019) Empirical analysis of network construction determinants of Azul Airlines. Working Paper. Center for Airline Economics - CAE. Disponível em www.nectar.ita.br/pesquisas.

Sinclair, R. A. (1995). An empirical model of entry and exit in airline markets. Review of Industrial Organization, 10(5), 541-557.